\pdfoutput=1

\documentclass{article}
\newcommand\tab[1][1cm]{\hspace*{#1}}

\usepackage[preprint]{neurips_2019}

\usepackage[utf8]{inputenc} 
\usepackage[T1]{fontenc}    
\usepackage{hyperref}       
\usepackage{url}            
\usepackage{booktabs}       
\usepackage{amsfonts}       
\usepackage{nicefrac}       
\usepackage{microtype}      
\usepackage{siunitx}
\usepackage[ruled,vlined]{algorithm2e}

\title{Adding Location and Global context to the Google/Apple Exposure Notification Bluetooth API}

\author{
  Ramesh Raskar \\
  MIT Media Lab\\
  Cambridge, MA 02139 \\
  \And
  Abhishek Singh \\
  MIT Media Lab\\
  Cambridge, MA 02139 \\
  \And
  Sam Zimmerman \\
  COVID Safe Paths\\
  Cambridge, MA 02139 \\
  \And
  Shrikant Kanaparti \\
  COVID Safe Paths\\
  Cambridge, MA 02139 \\
}

\begin{document}

    \maketitle
    \begin{abstract}
        Contact tracing requires a strong understanding of the context of a user, and location with other sensory data could provide a context for any infection encounter. Although Bluetooth technology gives a good insight into the proximity aspect of an encounter, it does not provide any location context related to it which helps taking better decisions. Using the ideas presented in this paper, one shall be able to obtain this valuable information which could address the problem of false positive and false negative to a certain extent. All of this within the purview of Google/Apple Exposure Notification (GAEN) specification, while preserving complete user privacy. There are four ways of propagating context between any two users. Two such methods allow private location logging, without revealing the location history within an app. The other two are encryption-based methods. The first encryption method is a variant of Apple's FindMy protocol, that allows nearby Apple devices to capture the GPS location of a lost Apple device. The second encryption is a minor modification of the existing GAEN protocol, so that global context is available to a healthy phone only when it is exposed - this is a better option comparatively. It will still be the role of Public Health smartphone app to decide, on how to use the location-time context, to build a full-fledged contact tracing and public health solution. Lastly, we highlight the benefits and potential privacy issues with each of these context propagation methods proposed here.
    \end{abstract}

    \section{Motivation}
    Currently, exposure notification obtained from GAEN, only reports the day of an exposure, but does not give any details such as their location or time. We believe that the context of location and time become critical for (i) a user to self-assess their exposure (e.g. if they were wearing a mask, or maintaining social distance at that moment) and inform about the same to those who were around them, or in case they were not carrying their smartphone at all at that point in time, (ii) improve user’s trust in the system to reject false-positives (e.g. if they had picked up BLE signal from behind a wall), (iii) help public health officials to perform contact tracing operations more accurately, also requires knowledge of the context (e.g. to request all those attending a wedding to self-isolate at their homes, if an infected person was observed to be at that place, and for a long duration). Lack of context and the lack of an agency can lead to irrational behavior, and civil unrest as explained in our document \href{https://docs.google.com/document/d/1fKmDvc3ip9Br9veyL8XYBFCjn0HBWHsxDF9vQKw-Q5s/edit}{Contact Tracing: Holistic Solution beyond Bluetooth}.\\
    \paragraph{What is Global Context} The idea behind global context is to deliver information which would improve exposure notification's accuracy or improve the actions followed by the exposure notification. One important component of the global context is GPS based location (which we also use interchangeably in this paper), in addition, it can have time, phone orientation, motion sensor (in hand or pocket), and any multipath (indoor vs outdoor) information.
    \\
    A user's behaviour and the level of precautionary measures they may need to take may vary depending on their specific circumstances. The risk from an exposure to a carrier also varies in different situations. Thus, the need for spatio-temporal context becomes imperative in contact tracing. Beyond location, there are many other context factors which can be utilized such as - Barometer reading, Ambient light sensor reading, magnetometer reading, gyroscope reading and etc. These sensor reading can unlock the potential of improving the false positive and false negative issues inherently associated with the bluetooth technology~\cite{leith2020coronavirus}
    
    Our goal is to enable location-time context to be delivered to the user during exposure notification. Here are a few possibilities and challenges.
    
    \begin{itemize}
        \item \emph{Log location by the same app} \\ An app could be rejected for storing time or location along with its Bluetooth packets, which makes it non-compliant under the current specification of Google/Apple Exposure Notification (GAEN) API.
        \item \emph{Use a secondary app} \\ One can store the GPS trails or any other similar context along with their timestamp using another app, which creates the issue of communication between these two apps, and also prevents mass adoption, as only a small percentage would install and run both these apps in parallel.
        \item \emph{Reuse ENIN information} \\ The GAEN approved app can estimate timestamp from ENIN, which is windowed at every ten minutes and uses a tolerance window of $\pm$2 hours during the diagnosis key matching. So, one way of achieving this is to use the existing protocol, and use ENIN timestamp from the Bluetooth payload as a primary key between the two databases, and look up global context for intersecting code using another app.
    \end{itemize}
    
    Another important aspect of contact tracing that has been completely overlooked by a majority of existing Bluetooth-based approaches is the lack of assistance for the Policy Makers, Epidemiologists, City Officials, Response Teams, etc. With the availability of location context in a contact tracing app, it would enable the covid positive citizens, to give their consent more effectively, while providing data to the aforementioned decision-makers; and this could shape the policy significantly, to a level where the outcomes could drastically change, if right interventions are performed.
    
    \section{Related Work}
    A significant amount of work has been done around contact tracing in the last six months of the COVID-19 outbreak. ~\cite{raskar2020covid19} and ~\cite{2020covid19} provide a comprehensive survey of the contact tracing ecosystem across different parameters. In this work we are mainly interested in the bluetooth based protocol built by Google~\cite{gaengooglewebsite} and Apple~\cite{gaenapplewebsite} which circumvents the system level problem known with the bluetooth in background for the phones and hence has been adopted widely across many countries and states. There are few other protocols which also utilize bluetooth for proximity sensing ~\cite{chan2020pact,epione}, however, they suffer with the same issue of the hardware compatibility which stops any app from running in the background in iOS. Goal of this work is to ensure privacy and ethical aspects~\cite{raskar2020apps} while providing a user some context about the exposure which will make the contact tracing more effective.
    
    \section{SafePaths protocol to combine Global context + GAEN}
    We propose a protocol that stays completely within the boundaries of GAEN specification for both the bluetooth~\cite{gaenble} as well as cryptography~\cite{gaencrypto}, based on the idea that Global context information is made available to a healthy phone, only when it is exposed to an infected person around. If Alice is healthy and comes in proximity to Bob, who later was diagnosed Covid+, then Alice will be able to see the location (and time) of that encounter, but rest of her location history shall remain invisible or encrypted - based on minor modification to GAEN, as suggested here. We also propose three other ideas, that do not require any modifications to the GAEN protocol, but rather require Google/Apple to allow apps to access the global context privately on the device and to run independent servers.
    
    Using BLE for proximity, and GPS, sensory data for context exclusively, the 4 privacy preserving solutions we propose that provide location-time context can be summarized as follows:
    \begin{enumerate}
        \item Global context logs stay on device app, does not leave the phone, and no visualization
        \item Global context + Time blurred and logged on device, data does not leave the phone, no visualization
        \item FindMy variant: Encrypt RPI
        \item GAEN variant: Encrypt your own Global context with DailyKey, and Broadcast it over BLE
    \end{enumerate}
    
    Please refer to this document for \href{https://covid19-static.cdn-apple.com/applications/covid19/current/static/contact-tracing/pdf/ExposureNotification-CryptographySpecificationv1.1.pdf}{GAEN terminology} that has been used here.\\
    Unless stated specifically, a Healthy person is $X$ (Alice), and a person who gets diagnosed of Covid is $Y$ (Bob).
    
    \emph{BT} :  Bluetooth \\
    \emph{BLE} :  Bluetooth Low Energy \\
    \emph{RPI} :  Rolling Proximity Identifiers, are privacy-preserving identifiers that are broadcast in Bluetooth payload \\
    \emph{ENIN} :  Exposure Number Interval Number (Index of a 10 min window, 144 such windows per day) \\
    \emph{DailyKey} :  Diagnosis Keys, a subset of which becomes Temporary Exposure Keys (previously known as Daily Tracing Keys)
    
    \begin{enumerate}
        \item \textbf{GPS logs stay on device app, data does not leave the phone, and cannot be visualized} \\ \\
        Approach 1 - based on direct RPI indexing: Every user logs their location, indexed with RPI. When exposure notification arrives, it matches the RPI with the logged location. We store the location only if there is RPI (which means another BT user was encountered). In this case the database of GPS is indexed with RPI. \\
        Approach 2 - based on calculated ENIN: The 144 RPIs generated from the DailyKey provides a match with one of the RPIs (10 minute window) so app has access to the timestamp. From that timestamp, the app can recover the GPS location for that time. In this case the database of GPS is indexed with time. \\
        No direct visualization of location log is made available to the user, to prevent unauthorized persons (nosy employers, abusive spouses or border agents) from seeing it. \\ \\
        \emph{Benefits}
        \begin{itemize}
            \item Location data stays local and casual unauthorized reader cannot see it (e.g. nosy employer, abusive spouse).
        \end{itemize}
        \emph{Issues}
        \begin{itemize}
            \item The app could get rejected, because it stores location log, which is against the GAEN guidelines.
            \item Sophisticated hackers can reverse engineer the exact location from logs. \\
        \end{itemize}

        \item \textbf{BLE for proximity, GPS for context, GPS is blurred for privacy, data does not leave the phone, and cannot be visualized}
        \begin{center}
            \emph{This is our recommended protocol if GAEN and GPS APIs can co-exist.}
        \end{center}
        Same as solution 1, but the location-time is quantized in space-time, in order to blur it before storing locally on the phone. For example, based on the local population density, the blur can be about 200m in a city or 1km in the suburbs. With the help of user's own memory, the user can figure out where the exposure encounter could have happened. \\ \\
        \emph{Benefits}
        \begin{itemize}
            \item A nosy employer cannot see specific location data.
            \item A sophisticated hacker who can reverse engineer, cannot get the exact location because it has been quantized.
        \end{itemize}
        \emph{Issues}
        \begin{itemize}
            \item Context could be reduced significantly, and hence this method appears to bring the classic dogma of utility vs privacy tradeoff.
            \item Context could also be wrong in some cases.
            \item Possibility of on-the-fly attacks, when an attacker has prior knowledge of the precise location. \\
        \end{itemize}

        \item \textbf{GPS is encrypted with the RPI of an infected person} \\
        Borrowing the concept of “upload what you heard” described in \href{https://support.apple.com/guide/security/end-to-end-encryption-sec60fd770ba/web}{FindMy protocol}. Every user encrypts their own GPS value, using public key of the user nearby and stores them locally. The healthy user phone then downloads the DailyKey and computes the list of RPIs that are related to it, and checks for the intersections if any with the list of heard RPIs using the existing GAEN API. Using which the corresponding encrypted location data is decrypted using their own private key. Note that both UUID$_x$ and PublicKey$_x$ keep rotating at every 15 minutes interval. Rotating PublicKey$_x$ while keeping the same PrivateKey$_x$ in an efficient manner is non-trivial, hence we propose to use the already built-in FindMy device protocol by Apple. \\ \\
        \begin{algorithm}[H]
        \caption{Encryption of context with infected person's RPI}
        
            $X$ is a healthy person and $Y$ is infected\;
            $X$ emits (UUID$_x$, PublicKey$_x$)\;
            $Y$ receives (UUID$_x$, PublicKey$_x$)\;
            $Y$ encrypts its own GPS as $D =$ Encrypt(PublicKey$_x,$ GPS$_y$) at this point\;
            stores it locally (GPS$_x$ == GPS$_y$)\;
            $Y$ gets identified as infected\;
            Y uploads the data it had stored into a diagnosis server (UUID$_x$, $D_x$)\;
            
            $X$ pulls data from the diagnosis server\;
            $X$ performs a comparison, and finds a match with its UUID$_x$\;
            $X$ uses its own PrivateKey$_x$ obtains:
            GPS$_x$ = Decrypt(PrivateKey$_x$, $D_x$)
        \end{algorithm}
        
        \emph{Challenges}
        \begin{itemize}
            \item Scalability, as this would need a lot of public keys to be encrypted
            \item A sophisticated attacker could reverse engineer the key $X$ \\
        \end{itemize}

        \item \textbf{GPS in existing GAEN protocol} \\
        “Upload what you broadcasted” as in GAEN protocol (as well as MIT PACT). In the following proposals we extend the existing GAEN scheme to allow context sharing. We propose two ways for this: one is based on \emph{Asymmetric key} infrastructure, while the other is based on \emph{Symmetric key} infrastructure. \\
        Our goal is to use the new Associated Encrypted Metadata (AEM) supported in GAEN. AEM currently is for the purpose of sharing BLE signal strength. We assume that we can append GPS value into this metadata, which gets broadcast over the BLE. \\
        Using this protocol, \emph{Alice can recover the GPS location of her contact (and hence her own location) only for locations for which she received the Exposure Notification}. She does not have access to the rest of her location history or the infected person’s location history.
        \begin{enumerate}
            \item \textbf{Using Asymmetric key encryption - Algo Steps} \\
            Every device has a Rotating Public Key - RPublicKey, and a single PrivateKey. This RPublicKey is different from the RPI, that is used in GAEN. We aim to encrypt the GPS with this RPublicKey, and the healthy user can then recover the encrypted GPS, using the corresponding PrivateKey downloaded from the server. \\ RPI = Rolling Proximity Identifiers (broadcast over BLE as in GAEN), \\ RPublicKey = Rotating Public Key (RPublicKey and a single PrivateKey) \\ \\
            \begin{algorithm}[H]
                \caption{Asymmetric key encryption based GAEN context encoding}
                Infected phone $Y$ emits \\ 
                \tab ( RPI$_y$, RPublicKey, \emph{Encrypt}(RPublicKey, GPS$_y$) )\\
                Healthy phone $X$, over BLE, receives this \\
                \tab ( RPI$_y$, RPublicKey, \emph{Encrypt}(RPublicKey, GPS$_y$) )
                \\ and stores it locally\\
                If infected, $Y$ uploads (DailyKey$_y$, PrivateKey$_y$)\\
                With GAEN, Healthy phone $X$, downloads (DailyKey$_y$, PrivateKey$_y$)
                $X$ from DailyKey$_y$, generates \{RPI$_y$\}\\
                $X$ finds the corresponding entry \\
                \tab ( RPI$_y$, RPublicKey, and \emph{Encrypt}(RPublicKey, GPS$_y$) )\\
                $X$ obtains GPS$_y$ by \\
                \tab \emph{Decrypt}( PrivateKey$_y$, \emph{Encrypt}(RPublicKey, GPS$_y$) ) \\
            \end{algorithm}
            
            \item \textbf{Using Symmetric key encryption - Algo Steps}
            \begin{center}
                \emph{This is our recommended algorithm with minimal change to GAEN}
            \end{center}
            This approach does not change anything about the uploading of DailyKey. It changes the BLE payload to include GPS encrypted with DailyKey. The \emph{Encrypt} method could be AES (key size can be adjusted to use standard 128 or 256 bit size). \\ \\
            \begin{algorithm}[H]
                \caption{Symmetric key encryption based GAEN context encoding}
                Infected phone $Y$ emits \\
                \tab ( RPI$_y$, \emph{Encrypt}(DailyKey$_y$, GPS$_y$) )\\
                Healthy phone $X$, over BLE receives the same, and stores it locally \\
                \tab ( RPI$_y$, \emph{Encrypt}(DailyKey$_y$, GPS$_y$) )
                \\Once diagnosed as infected, phone $Y$ uploads its (DailyKey$_y$)
                \\ As part of GAEN, the Healthy phone $X$, downloads (DailyKey$_y$)
                \\ From DailyKey$_y$, phone $X$ generates its RPI$_y$
                \\ $X$ Identifies the corresponding entry (RPI$_y$, \emph{Encrypt}(DailyKey$_y$, GPS$_y$))
                \\ $X$ Extracts GPS$_y$ by \\
                \tab \emph{Decrypt}( DailyKey$_y$, \emph{Encrypt}(DailyKey$_y$, GPS$_y$) ) \\
            \end{algorithm}
            
            \item \textbf{Using Symmetric key encryption and \emph{consent after infection}} \\
            We allow a consent mechanism after the broadcast. The user may have consented to broadcast BLE Ids, and also shared their encrypted GPS. But the infected user may change their opinion to share only the BLE, and not allow the healthy user to decrypt their GPS location. User keeps a ‘ConsentSecret’ code, and this is appended to the DailyKey\footnote{alternatively one also could perform an Exclusive-OR (XOR) operation of DailyKey and ConsentSecret}. The RPI is derived from DailyKey, and hence BLE decoding is not impacted. But GPS is encrypted with the key that appends DailyKey$_y$ with ConsentSecret. If the inflected user refuses to upload the ConsentSecret, the broadcast encrypted GPS coordinates cannot be recovered. Rest of the protocol is the same as 4(b) above. \\ \\
            \begin{algorithm}[H]
                \caption{Symmetric key encryption based GAEN context encoding}
                Infected phone $Y$ emits \\
                \tab ( RPI$_y$, \emph{Encrypt}(DailyKey$_y$ $||$ ConsentSecret$_y$, GPS$_y$) ) \\ 
                ($||$ refers to the concatenation operator, for cryptographic reasons, it is advised to use XOR instead)
                \\Healthy phone $X$, over BLE, receives the same, and stores it locally
                \\ ( RPI$_y$, \emph{Encrypt}(DailyKey$_y$ $||$ ConsentSecret$_y$, GPS$_y$) )
                \\ Once diagnosed as infected, they shall upload their (DailyKey, ConsentSecret), where ‘ConsentSecret’ is a unique secret key for allowing consent to be provided for decrypting location context, or just provide exposure notification\footnote{For brevity we are keeping ConsentSecret here as a non-rotating unique secret key, but this can be changed in the upcoming version of the draft}.
                \\ With GAEN, Healthy phone $X$, downloads (DailyKey$_y$, ConsentSecret$_y$)
                \\ From DailyKey$_y$, reconstructs RPI\_y
                \\ Find the corresponding entry \\ 
                \tab ( RPI$_y$, \emph{Encrypt}(DailyKey$_y$ $||$ ConsentSecret$_y$, GPS$_y$) )
                \\ Extracts GPS$_y$ by \\ \tab \emph{Decrypt}( DailyKey$_y$ $||$ ConsentSecret$_y$, \\ \tab[3cm] \emph{Encrypt}(DailyKey$_y$ $||$ ConsentSecret$_y$, GPS$_y$) ) \\
            \end{algorithm}
            
            \item \textbf{Encrypted and Blurred GPS and post-infection Consent}
            \begin{center}
                \emph{This is our recommended algorithm if GPS is supported only in BLE payload.}
            \end{center}
            This does not change anything about the uploading of DailyKey. It rather changes the BLE payload to include a Blurred GPS encrypted with DailyKey. It allows infected user to change the Consent as frequently as DailyKey is changed. Currently, DailyKey changes per day, but in the future, they may change more frequently as in PACT, allowing more fine grained Consent by the infected user. \\ \\
            \begin{algorithm}[H]
                \caption{Symmetric key encryption based GAEN context encoding with blurred location}
                Infected phone Y emits \\ 
                \tab ( RPI$_y$, \emph{Encrypt}(DailyKey$_y$ $||$ ConsentSecret$_y$, FGPS$_y$) ) \\
                ($||$ refers to the concatenation operator) and FGPS$_y$ is quantized GPS. Quantization based blurring of location can be fixed, or vary from 100’s of meters to 1km depending on the population density around that location.
                \\ Healthy phone $X$, over BLE, receives the same, and stores it locally \\
                \tab ( RPI$_y$, \emph{Encrypt}(DailyKey$_y$ $||$ ConsentSecret$_y$, FGPS$_y$) )
                \\ Once diagnosed as infected, they upload (DailyKey, ConsentSecret), where ‘ConsentSecret’ is a unique secret key for allowing consent to be provided for decrypting location context or just provide exposure notification.
                \\ With GAEN, Healthy phone $X$, downloads (DailyKey$_y$, ConsentSecret$_y$)
                \\ From DailyKey$_y$, reconstructs RPI$_y$
                \\ Find the corresponding entry \\
                \tab ( RPI$_y$, \emph{Encrypt}(DailyKey$_y$ $||$ ConsentSecret$_y$, FGPS$_y$) )
                \\ Extracts FGPS$_y$ by \\ \tab \emph{Decrypt}( DailyKey$_y$ $||$ ConsentSecret$_y$, \\ \tab[3cm] \emph{Encrypt}(DailyKey$_y$ $||$ ConsentSecret$_y$, FGPS$_y$) ) \\ Which is a blurred location that is enough to provide context for the user, but does not provide the exact location.
            \end{algorithm}
        \end{enumerate}
        
        \emph{Benefits}
        \begin{itemize}
            \item Only DailyKey is uploaded as in GAEN, so no change in upload protocol
            \item A minor change in the BLE payload
            \item GPS (encrypted) is available only to the proximate phone, so there is little or no risk to a non-proximate person
            \item GPS history is invisible
        \end{itemize}
        
        \emph{Challenges}
        \begin{itemize}
            \item BLE payload increases but GAEN payload has plenty of space
        \end{itemize}
    \end{enumerate}

    \section{Discussion}
    We consider three possible levels of attack (The three levels are not mutually exclusive): \\ \\
    \emph{On-the-fly attack} \\ 
    This scheme of attack is performed by an attacker acting either as a Healthy person or a future-infected person. They can potentially do two kind of attacks:
    \begin{itemize}
        \item Snoop on information.
        \item Spread wrong information through the Bluetooth transmission.
    \end{itemize}  
    \emph{Post processing attack}
    \begin{itemize}
        \item In this setting of attack, the attacker tries to make sense out of encrypted and unencrypted information available after collection on their phone or force someone else to show these information present on their phone.
    \end{itemize}
    \emph{Distributed multi party attack}
    \begin{itemize}
        \item In this scheme, multiple individuals align together to share their data with each other in a distributed way to attack the secrecy and privacy of individuals or groups of individuals.
    \end{itemize}
    There are different threat actors against whom protection is required: \\ \\
    \emph{Nosy person looking at stored GPS trails because visualization is easy}
    \begin{itemize}
        \item Can force the user to open the app and show any data visible on the screen.
    \end{itemize}
    \emph{Hacker looking at stored GPS trails if available in raw format somewhere in the app}
    \begin{itemize}
        \item Can reverse engineer their own app to inject code on top of APIs (This is only possible by jailbreaking the iOS and rooting the Android OS).
        \item Can perform packet captures and snooping.
        \item Can not force the user to open app and share data with the attacker.
    \end{itemize}
    \emph{State actors reverse engineering information using poorly encrypted trails (using side channel)}
    \begin{itemize}
        \item Combines the capabilities of the above actors and in addition can leverage multiple sources and supercomputing capabilities for cryptanalysis.
    \end{itemize}
    From the GPS location, for an added context, the user may need to call a reverse geocoding API to find the street address or name of the business there. If performed naively, this API call will leak user location. The easiest way to resolve would be to perform regional map caching but this approach is beyond the scope of this document.

    \section{Conclusion}
    In this proposal we have outlined several ways of allowing context-enabled contact tracing. We believe the contextual information and time will allow citizens to take informed decisions and reduce panic.
    
    In the spirit of respecting privacy, allowing consent, and delivering context within the Exposure Notification service, we advocate for the adoption of Proposal 4(b).
    
    \section{Acknowledgement}
    We would like to thank Mikhail Dmitrienko for helping with the writing of the document.
    
\bibliography{main}

\begin{thebibliography}{10}
\providecommand{\natexlab}[1]{#1}
\providecommand{\url}[1]{\texttt{#1}}
\expandafter\ifx\csname urlstyle\endcsname\relax
  \providecommand{\doi}[1]{doi: #1}\else
  \providecommand{\doi}{doi: \begingroup \urlstyle{rm}\Url}\fi

\bibitem[gae(2020{\natexlab{a}})]{gaenapplewebsite}
\emph{Exposure Notification reference by Apple}, 2020{\natexlab{a}}.
\newblock URL \url{https://www.apple.com/covid19/contacttracing}.

\bibitem[gae(2020{\natexlab{b}})]{gaenble}
\emph{Exposure Notification, Bluetooth Specification}, 2020{\natexlab{b}}.
\newblock URL
  \url{https://covid19-static.cdn-apple.com/applications/covid19/current/static/contact-tracing/pdf/ExposureNotification-BluetoothSpecificationv1.2.pdf}.

\bibitem[gae(2020{\natexlab{c}})]{gaencrypto}
\emph{Exposure Notification, Cryptography Specification}, 2020{\natexlab{c}}.
\newblock URL
  \url{https://blog.google/documents/69/Exposure_Notification_-_Cryptography_Specification_v1.2.1.pdf}.

\bibitem[gae(2020{\natexlab{d}})]{gaengooglewebsite}
\emph{Exposure Notification reference by Google}, 2020{\natexlab{d}}.
\newblock URL \url{https://www.google.com/covid19/exposurenotifications/}.

\bibitem[Chan et~al.(2020)Chan, Foster, Gollakota, Horvitz, Jaeger, Kakade,
  Kohno, Langford, Larson, Sharma, Singanamalla, Sunshine, and
  Tessaro]{chan2020pact}
Chan, J., Foster, D., Gollakota, S., Horvitz, E., Jaeger, J., Kakade, S.,
  Kohno, T., Langford, J., Larson, J., Sharma, P., Singanamalla, S., Sunshine,
  J., and Tessaro, S.
\newblock Pact: Privacy sensitive protocols and mechanisms for mobile contact
  tracing, 2020.

\bibitem[Leith \& Farrell(2020)Leith and Farrell]{leith2020coronavirus}
Leith, D.~J. and Farrell, S.
\newblock Coronavirus contact tracing: Evaluating the potential of using
  bluetooth received signal strength for proximity detection, 2020.

\bibitem[Li \& Guo(2020)Li and Guo]{2020covid19}
Li, J. and Guo, X.
\newblock Covid-19 contact-tracing apps: a survey on the global deployment and
  challenges, 2020.

\bibitem[Raskar et~al.(2020{\natexlab{a}})Raskar, Nadeau, Werner, Barbar,
  Mehra, Harp, Leopoldseder, Wilson, Flakoll, Vepakomma, Pahwa, Beaudry,
  Flores, Popielarz, Bhatia, Nuzzo, Gee, Summet, Surati, Khastgir, Benedetti,
  Vilcans, Leis, and Louisy]{raskar2020covid19}
Raskar, R., Nadeau, G., Werner, J., Barbar, R., Mehra, A., Harp, G.,
  Leopoldseder, M., Wilson, B., Flakoll, D., Vepakomma, P., Pahwa, D., Beaudry,
  R., Flores, E., Popielarz, M., Bhatia, A., Nuzzo, A., Gee, M., Summet, J.,
  Surati, R., Khastgir, B., Benedetti, F.~M., Vilcans, K., Leis, S., and
  Louisy, K.
\newblock Covid-19 contact-tracing mobile apps: Evaluation and assessment for
  decision makers, 2020{\natexlab{a}}.

\bibitem[Raskar et~al.(2020{\natexlab{b}})Raskar, Schunemann, Barbar, Vilcans,
  Gray, Vepakomma, Kapa, Nuzzo, Gupta, Berke, Greenwood, Keegan, Kanaparti,
  Beaudry, Stansbury, Arcila, Kanaparti, Pamplona, Benedetti, Clough, Das,
  Jain, Louisy, Nadeau, Pamplona, Penrod, Rajaee, Singh, Storm, and
  Werner]{raskar2020apps}
Raskar, R., Schunemann, I., Barbar, R., Vilcans, K., Gray, J., Vepakomma, P.,
  Kapa, S., Nuzzo, A., Gupta, R., Berke, A., Greenwood, D., Keegan, C.,
  Kanaparti, S., Beaudry, R., Stansbury, D., Arcila, B.~B., Kanaparti, R.,
  Pamplona, V., Benedetti, F.~M., Clough, A., Das, R., Jain, K., Louisy, K.,
  Nadeau, G., Pamplona, V., Penrod, S., Rajaee, Y., Singh, A., Storm, G., and
  Werner, J.
\newblock Apps gone rogue: Maintaining personal privacy in an epidemic,
  2020{\natexlab{b}}.

\bibitem[Trieu et~al.(2020)Trieu, Shehata, Saxena, Shokri, and Song]{epione}
Trieu, N., Shehata, K., Saxena, P., Shokri, R., and Song, D.
\newblock Epione: Lightweight contact tracing with strong privacy, 2020.

\end{thebibliography}
\bibliographystyle{icml2019}
\end{document}